\documentclass[aps,pra,showpacs,twocolumn,nofootinbib]{revtex4-2}
\usepackage{bm}
\usepackage{amsmath}
\usepackage{amssymb}
\usepackage{slashed}
\usepackage{float}
\allowdisplaybreaks
\usepackage{subcaption}
\usepackage{dcolumn}
\usepackage{graphicx}

\usepackage{physics}
\usepackage{nicefrac}
\newcolumntype{w}[1]{D{.}{.}{#1}}
\newcommand{\Za}{Z\alpha}
\newcommand{\vare}{\varepsilon}

\newcommand{\bfx}{{\vec{x}}}
\newcommand{\bfz}{{\vec{z}}}

\begin{document}

\title{Two-loop electron self-energy for low nuclear charges}

\author{V.~A. Yerokhin}
\email{vladimir.yerokhin@mpi-hd.mpg.de}
\affiliation{Max~Planck~Institute for Nuclear Physics, Saupfercheckweg~1, D~69117 Heidelberg, Germany}

\author{Z. Harman}
\affiliation{Max~Planck~Institute for Nuclear Physics, Saupfercheckweg~1, D~69117 Heidelberg, Germany}

\author{C.~H. Keitel}
\affiliation{Max~Planck~Institute for Nuclear Physics, Saupfercheckweg~1, D~69117 Heidelberg, Germany}

\begin{abstract}

Calculations of the two-loop electron self-energy for the $1S$ Lamb shift are reported, performed
to all orders in the nuclear binding strength parameter
$\Za$ (where $Z$ is the nuclear charge number and $\alpha$ is the fine structure constant).
Our approach allows calculations to be extended to nuclear charges lower than previously possible
and improves the numerical accuracy by more than an order of magnitude.
Extrapolation of our all-order results to hydrogen yields
a result twice as precise as the previously accepted value
[E.~Tiesinga {\em et al.} Rev. Mod. Phys. {\bf 93}, 025010 (2021)],
differing from it by $2.8$ standard deviations.
The resulting shift in the theoretical prediction for the $1S$-$2S$ transition frequency
in hydrogen decreases the value of the Rydberg constant by one standard deviation.

\end{abstract}

\maketitle

The Rydberg constant is
one of the most precisely known physical constants. It is also
the cornerstone of atomic spectroscopy, as all atomic transition frequencies are proportional to it. Any theoretical predictions of atomic transitions must be multiplied by the Rydberg constant to be compared with experimental data. The most accurate determination of the Rydberg constant today involves combining the experimental $1S$-$2S$ transition frequency of hydrogen \cite{matveev:13}, the proton charge radius extracted from muonic hydrogen spectroscopy
\cite{antognini:13,pachucki:24:rmp}, and the theoretical prediction for the $1S$ and $2S$ Lamb shift \cite{tiesinga:21:codata18}. Among these three components, the uncertainty in the theoretical prediction is the largest and dominates the overall uncertainty of the Rydberg constant.

\begin{figure}
\centerline{
\resizebox{\columnwidth}{!}{%
  \includegraphics{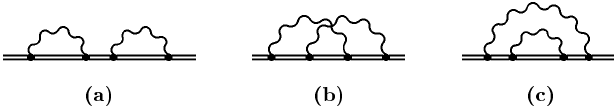}
}}
 \caption{
Feynman diagrams representing the two-loop electron self-energy. The double line denotes the electron in the
presence of the binding nuclear field; the wavy line denotes the exchange of
a virtual photon.
\label{fig:sese}}
\end{figure}

The two-loop electron self-energy (SESE, Fig.~\ref{fig:sese}) is arguably the most
problematic QED effect in the hydrogen Lamb shift,
which repeatedly pushed theoretical predictions beyond their error margins over recent decades.
In 1994,
Pachucki's calculation of the SESE contribution of order $m\alpha^2(\Za)^5$ \cite{pachucki:94}
revealed a significant effect that resolved\footnote{As
Steven Weinberg remarked on this occasion, ``So apparently
quantum electrodynamics wins again''\cite{weinberg:book}.}
a disturbing discrepancy \cite{weitz:94} between theory and experiment
existing at that time.
Seven years later, another calculation by Pachucki
\cite{pachucki:01:pra}  identified a large SESE contribution of order
$m\alpha^2(\Za)^6\ln(\Za)^{-2}$,
once again correcting the previous theoretical prediction \cite{eides:01}.
Finally, in 2005 and 2009, nonperturbative (in $\Za$) calculations of the SESE
correction were conducted for hydrogen-like ions, but only for $Z\ge 10$
\cite{yerokhin:05:sese,yerokhin:09:sese}.
These calculations accounted for higher orders in $\Za$ that are typically
inaccessible by other means.
Their results were successfully  validated by experiments in the high-$Z$ region
\cite{beiersdorfer:98,beiersdorfer:05,yerokhin:06:prl}, but required
extrapolation to offer insights for hydrogen.

Extrapolations of the nonperturbative results
towards $Z = 1$ reported in Refs.~\cite{yerokhin:05:sese,yerokhin:09:sese}
revealed some tension with predictions of the $\Za$ expansion
\cite{pachucki:03:prl}. However,
it was argued \cite{karshenboim:19:sese,yerokhin:18:hydr} that this could be
plausibly attributed to higher-order $\Za$ contributions. So,
the presently accepted value for hydrogen \cite{tiesinga:21:codata18}
was obtained by assuming the consistency of the $\Za$-expansion
and the all-order results. The optimistic error obtained in this manner
nevertheless
constitutes one of the two primary theoretical uncertainties in the hydrogen Lamb shift.
In this Letter we demonstrate that the enhanced precision of numerical calculations
challenges the assumed consistency between the nonperturbative and $\Za$-expansion results.
Consequently, we find that the theoretical prediction for the hydrogen Lamb shift needs to be
once again reexamined, with implications for the Rydberg constant.

Previous nonperturbative calculations of the SESE correction
\cite{yerokhin:05:sese,yerokhin:09:sese} were performed for hydrogen-like ions
with $Z\ge 10$. Technical difficulties of these computations increased rapidly as
$Z$ decreased, becoming overwhelming for the lowest nuclear charges.
This was partly due to numerical cancellations that
scaled as $1/Z^2$, but the main limiting factor was
the partial-wave expansion of the electron propagators.
This expansion is unbounded
and needs to be truncated in actual computations, with the number of
terms growing as $(2L)^3$ with increase of the cutoff parameter
$L$.
These issues prevented any further progress in improving the calculations
\cite{yerokhin:05:sese,yerokhin:09:sese} in the low-$Z$ region.

In this Letter, we utilize the advanced subtraction scheme suggested in Ref.~\cite{sapirstein:23} for the one-loop self-energy, which significantly improved the convergence of the partial-wave expansion in the one-loop case,
see also Ref.~\cite{malyshev:24}. We generalize this approach to the two-loop case and demonstrate that it leads to a drastic improvement in the partial-wave convergence,
thereby enabling a break-through in the achievable
accuracy of the two-loop computations.

The two-loop self-energy correction is represented by three Feynman diagrams
shown in Fig.~\ref{fig:sese}. The diagram (a), known as the loop-after-loop
diagram, is significantly simpler to compute than the rest and was calculated
for low nuclear charges already in Refs.~\cite{mallampalli:98:prl,yerokhin:00:lalpra}. The
remaining two diagrams in Fig.~\ref{fig:sese} present the main
computational challenge. The diagram
in Fig.~\ref{fig:sese}(b) induces the so-called {\em overlapping} contribution, whose
formal expression is given by
\begin{align}  \label{eq5}
\Delta E_{O} &\, =  (2i\alpha)^2
   \int_{C_F} d\omega_1d\omega_2
   \int d\bfx_1 d\bfx_2 d\bfx_3 d\bfx_4\,
    D^{\mu\nu}(\omega_1,x_{13})\,
\nonumber \\ &  \times
    D^{\rho\sigma}(\omega_2,x_{24})\,
   {\psi}^{\dag}_a(\bfx_1)\, \alpha_{\mu}\, G(\vare_a-\omega_1,\bfx_1,\bfx_2)\,
           \alpha_{\rho}\,
\nonumber \\ &  \times
  G(\vare_a-\omega_1-\omega_2,\bfx_2,\bfx_3)\,\alpha_{\nu}\, G(\vare_a-\omega_2,\bfx_3,\bfx_4)\,
\nonumber \\ &  \times
         \alpha_{\sigma}
     \psi_a(\bfx_4)\,,
\end{align}
where
$D^{\mu\nu}(\omega,x)$ is the photon propagator,
$\psi_a(\bfx)$ is the reference-state wave function with the energy $\vare_a$,
$\alpha_i$ are the Dirac matrices,
$G(\vare,\bfx_i,\bfx_j)$ is the
Dirac-Coulomb Green function,
$x_{ij} = |\bfx_i-\bfx_j|$, and $C_F$ is
the Feynman integration contour.
The diagram
in Fig.~\ref{fig:sese}(c) gives rise to the {\em nested} contribution,
expressed as

\begin{align}  \label{eq4}
\Delta E_{N} &\, = (2i\alpha)^2 \int_{C_F} d\omega_1d\omega_2\,
   \int d\bfx_1 d\bfx_2 d\bfx_3 d\bfx_4\,
    D^{\mu\nu}(\omega_1,x_{14})\,
\nonumber \\ &  \times
    D^{\rho\sigma}(\omega_2,x_{23})\,
       {\psi}^{\dag}_a(\bfx_1)\, \alpha_{\mu}
          G(\vare_a-\omega_1,\bfx_1,\bfx_2) \, \alpha_{\rho}\,
\nonumber \\ &  \times
           G(\vare_a-\omega_1-\omega_2,\bfx_2,\bfx_3)\, \alpha_{\sigma}\,
          G(\vare_a-\omega_1,\bfx_3,\bfx_4)\,
\nonumber \\ & \times
          \alpha_{\nu}\, \psi_a(\bfx_4)\,.
\end{align}

\begin{figure}
\centerline{
\resizebox{\columnwidth}{!}{%
  \includegraphics{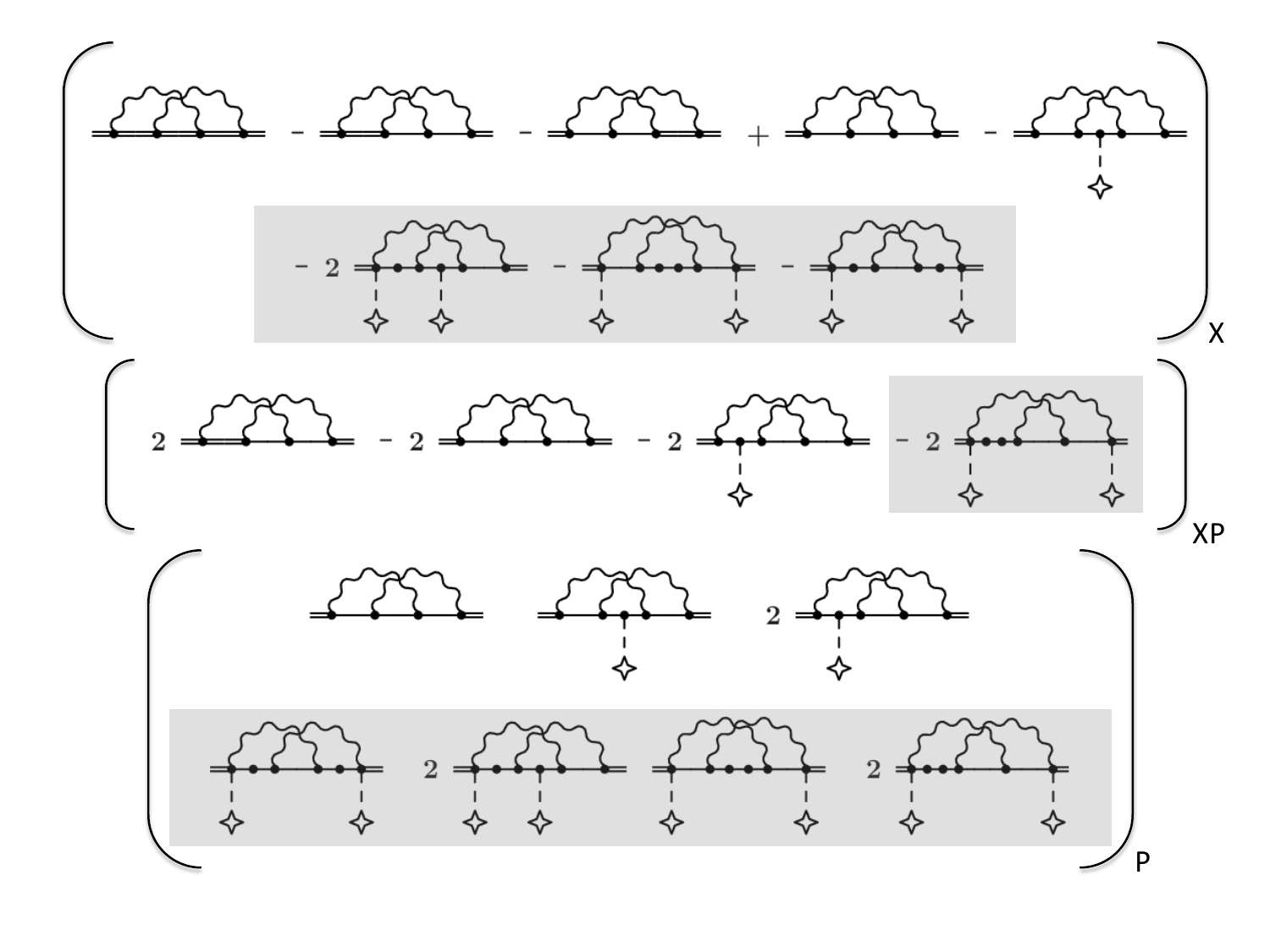}
}}
 \caption{
Two subtraction schemes used for calculations of the
overlapping diagram: the previous scheme (without shaded diagrams) and
the present scheme (with shaded
diagrams), see Supplementary Material \cite{suppl}.
The single solid line represents the free electron propagator,
while the dashed line terminated by a stylized cross indicates the
interaction with the binding nuclear field.
\label{fig:nested}}
\end{figure}

\begin{figure}
\centerline{
\resizebox{\columnwidth}{!}{%
  \includegraphics{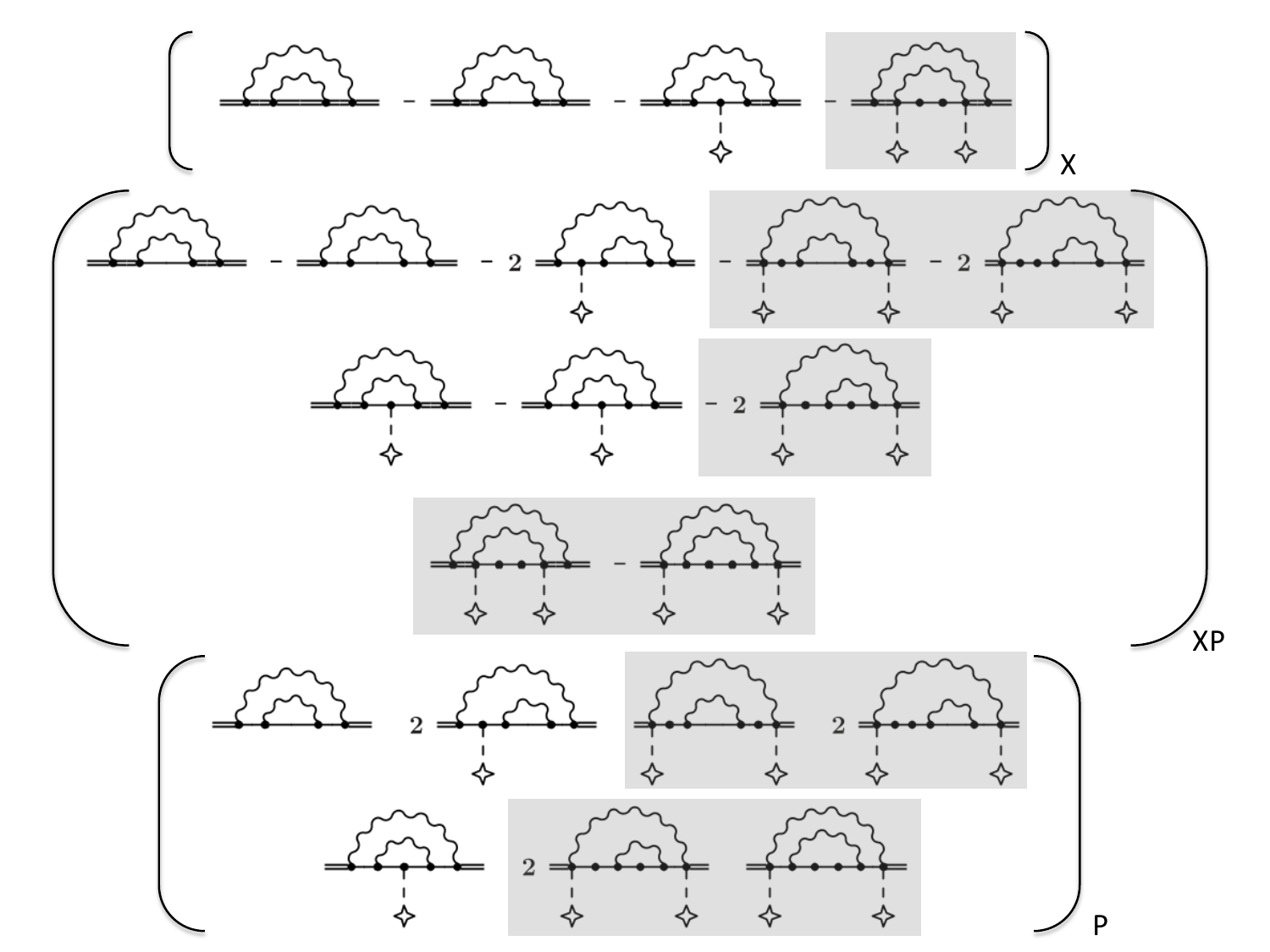}
}}
 \caption{Same as Fig.~\ref{fig:nested} for the
nested diagram. \label{fig:overlap}}
\end{figure}

The above formulas are formal expressions that require renormalization before any actual calculations can be performed. The renormalization scheme is based on subtracting and re-adding several first terms of the expansion of the bound-electron propagators in terms of interactions with the binding field. This procedure greatly increases the number of diagrams that need to be computed. The renormalization scheme was first formulated in Ref.~\cite{mallampalli:98:pra} and fully implemented in a series of our studies \cite{yerokhin:01:sese,yerokhin:03:prl,yerokhin:06:prl}. It is schematically depicted in Figs.~\ref{fig:nested} and \ref{fig:overlap}. The diagrams to be computed are divided into three classes: (i) those calculated in coordinate space, labeled "X," (ii) those calculated in the mixed momentum-coordinate representation, labeled "XP," and (iii) those computed in momentum space, labeled "P." Details of the calculation method are described in our previous investigations \cite{yerokhin:03:epjd,yerokhin:10:sese,yerokhin:18:sese}. The major limiting factor for the numerical accuracy achievable in this scheme is the convergence of the (double) partial-wave expansion of diagrams calculated in the coordinate space.

In this Letter, we improve upon the previous subtraction scheme to achieve better partial-wave convergence for the diagrams calculated in coordinate space and the mixed momentum-coordinate representation. Following the general idea of Ref.~\cite{sapirstein:23}, we subtract and re-add diagrams containing two explicit interactions with the binding nuclear field. There is some flexibility in choosing these subtraction diagrams since the Coulomb potentials can be "moved" to neighboring vertices without significantly affecting the behavior of the partial-wave expansion. This flexibility allows us to select subtraction diagrams that can also be computed in closed form (without any partial-wave expansion) in momentum space. The additional subtraction diagrams are depicted in Figs.~\ref{fig:nested} and \ref{fig:overlap}, indicated by the shaded background,
and described in Supplementary Material \cite{suppl}. As a result, we typically achieve an improvement in the convergence of partial-wave expansions by one or two orders of magnitude in terms of achievable numerical precision. The trade-off for this improvement is the significantly larger number of diagrams that need to be computed.

Numerical results of our calculation are presented in Table~\ref{tab:sese}, expressed
in terms of the dimensionless function $F(\Za)$, which is connected to the energy
shift $\Delta E$ by
\begin{align}
\Delta E = mc^2\, \frac{\alpha^2}{\pi^2} (\Za)^4\,F(\Za)\,.
\end{align}
The comparison
presented in the table shows that our new results are consistent with the previous
calculations \cite{yerokhin:09:sese,yerokhin:18:sese} and significantly improve upon
them. For example, at $Z = 10$, the numerical accuracy is improved by a factor of 30.
Additionally, we extend our computations to lower values of $Z$, down to $Z = 5$.
This improvement allows for a better extrapolation towards $Z = 1$
than was previously possible, as will be discussed below.
It should be noted that the implementation of the new subtraction scheme provided an additional cross-check of the numerical procedure used in our previous calculations. This is because the new subtraction diagrams are computed by two different methods: with partial-wave expansion and in the closed form.

\begin{table}[b]
    \caption{Two-loop self-energy correction for the $1s$ state of H-like ions.
    \label{tab:sese}}
\begin{ruledtabular}
    \begin{tabular}{l w{2.7} w{2.7} w{3.9}}
 \multicolumn{1}{c}{$Z$}   & \multicolumn{1}{c}{$F(\Za)$}
       & \multicolumn{1}{c}{$F(\Za)$, previous }
    & \multicolumn{1}{c}{$G_{60}(\Za)$} \\
\hline\\[-5pt]
  5   &         0.6665\,(70)   &       &         -99.21\,(524)    \\
  6   &         0.5474\,(36)   &       &        -100.51\,(190)    \\
  7   &         0.4374\,(28)   &       &        -100.28\,(107)    \\
  8   &         0.3342\,(15)   &       &         -99.74\,(43)     \\
  9   &         0.2385\,(13)   &       &         -98.66\,(29)     \\
 10   &         0.1487\,(12)   &  0.17\,(4)^a     &         -97.47\,(22)     \\
 12   &        -0.0146\,(11)   &  0.00\,(4)^a     &         -94.85\,(14)     \\
 14   &        -0.1592\,(10)   &       &         -92.08\,(10)     \\
 16   &        -0.2880\,(11)   &       &         -89.241\,(83)    \\
 18   &        -0.4043\,(10)   &       &         -86.459\,(58)    \\
 20   &        -0.5092\,(8)    & -0.501\,(5)^a      &         -83.682\,(36)    \\
 22   &        -0.6046\,(8)    &       &         -80.947\,(33)    \\
 26   &        -0.7709\,(7)    &       &         -75.577\,(20)    \\
 30   &        -0.9140\,(9)    & -0.912\,(3)^b      &         -70.434\,(18)    \\
 34   &        -1.0384\,(9)    &       &         -65.467\,(15)    \\
 40   &        -1.2014\,(8)    & -1.199\,(3)^b      &         -58.3741\,(91)   \\
 50   &        -1.4382\,(5)    & -1.438\,(2)^b      &         -47.4279\,(39)   \\
    \end{tabular}
\end{ruledtabular}
$^a$ Ref.~\cite{yerokhin:09:sese}; $^b$ Ref.~\cite{yerokhin:18:sese}.
\end{table}

In order to compare our numerical all-order results with those obtained
within the $\Za$ expansion, we need to separate out contributions of the
leading $\Za$-expansion terms.
The $\Za$ expansion of function $F(\Za)$ for an $S$ state has the form
\begin{align} \label{eq:1}
F(\Za)  = B_{40} + &\  (\Za)\,B_{50} + (\Za)^2\big[ B_{63}\,L^3
 \nonumber \\ &
+ B_{62}\,L^2 + B_{61}\,L +
 G_{60}(\Za)\big]\,,
\end{align}
where $L = \ln (\Za)^{-2}$, $G_{60}(\Za)$ is the remainder containing all higher-order terms
in $\Za$, and coefficients $B_{ij}$ can be found in Ref.~\cite{yerokhin:18:hydr}.
The leading coefficients $B_{40}$, $B_{50}$, and $B_{63}$ have been confirmed through independent calculations
and are well established, whereas the coefficients $B_{62}$ and $B_{61}$
derived by Pachucki \cite{pachucki:01:pra} have yet to be independently verified.
We now use the analytical results to extract the higher-order remainder
function $G_{60}(\Za)$ from our calculations (listed in Table~\ref{tab:sese})
and will return to the discussion of potential modifications to the analytical coefficients at a later stage.

For the extrapolation of the higher-order remainder $G_{60}(\Za)$, it is important to know the functional
form of its $\Za$ expansion. It was worked out in Ref.~\cite{karshenboim:19:sese}
and is given by
\begin{align}\label{eq:2}
 G_{60}(\Za) = &\ B_{60} + (\Za)\big[ B_{72} \, L^2+  B_{71} \, L + B_{70}\big]
  \nonumber \\ &
  + (\Za)^2 \big[ B_{84}\,L^4 + B_{83}\,L^3 + B_{82}\,L^2
    \nonumber \\ &
\hspace*{2cm}
    + B_{81}\,L
+ B_{80}\big] + \ldots\,.
\end{align}
The leading logarithmic coefficients in the above formula are known exactly, $B_{72} = -6.19408\ldots$
\cite{karshenboim:18} and $B_{84} = -7/27$ \cite{karshenboim:19:sese}. The coefficient $B_{60}$
was partially computed in Ref.~\cite{pachucki:03:prl}, with the result $B_{60} = -61.6\,(9.2)$, where
the uncertainty represents the estimation of unevaluated contributions. For other coefficients in
Eq.~(\ref{eq:2}), Ref.~\cite{karshenboim:19:sese} reported approximate constraints. In this Letter,
we will need the constraints for the logarithmic coefficients,
\begin{align}\label{eq:3}
&B_{71} = -12\pm40\,,\, B_{83} = 0\pm 3.2\,,\,
\nonumber \\ &
B_{82} = 0\pm 50\,, \, B_{81} = 0\pm 150\,.
\end{align}

\begin{figure}[t]
\centerline{
\resizebox{1.15\columnwidth}{!}{%
  \includegraphics{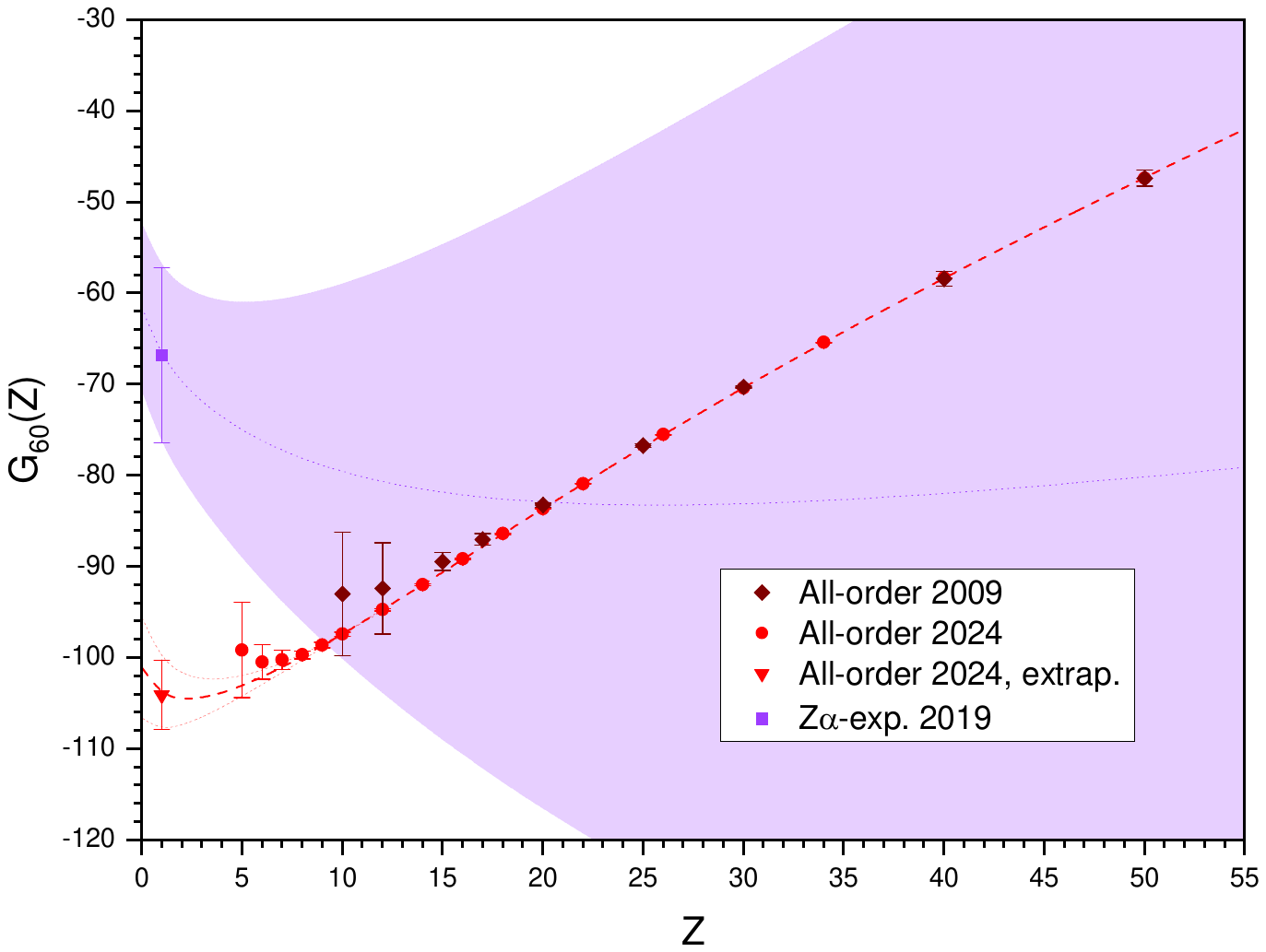}
}}
 \caption{
Two-loop self-energy higher-order function $G_{60}(Z)$ defined in Eq.~(\ref{eq:1}),
in the nonperturbative approach (the present calculations, red dots; the previous calculations
\cite{yerokhin:09:sese}, brown diamonds;
the fit of the present data, red dashed line and triangle),
alongside with the prediction of the
$\Za$-expansion approach
\cite{karshenboim:19:sese} (violet square and filled area).
\label{fig:ho}}
\end{figure}

Our nonperturbative results for the higher-order remainder $G_{60}(\Za)$ are plotted in Fig.~\ref{fig:ho}, alongside the previously reported values from Refs.~\cite{yerokhin:09:sese,yerokhin:18:sese}, and the $\Za$-expansion approximation from Ref.~\cite{karshenboim:19:sese}. We observe that the previous all-order results fall within the shaded region of the approximate $\Za$-expansion prediction, whereas the new results clearly lie outside of that region. This indicates that the results obtained from the two methods are now in disagreement. For this reason,
we will not use the analytical prediction for the $B_{60}$ coefficient together with our numerical results for extrapolation.
At the same time, we will have to rely on the analytical constraints for the logarithmic coefficients as given by Eq.~(\ref{eq:3}) in our fitting.
This is because the current numerical data cannot constrain the logarithmic coefficients $B_{71}$, $B_{83}$, etc.
The details of the fitting procedure are described in Supplementary Material \cite{suppl}.

The results of extrapolation of our numerical data are plotted in Fig.~\ref{fig:ho}.
The structure of the fitting curve at $Z \to 0$ is determined by the analytical values
of the $B_{72}$ and $B_{71}$ coefficients. For the
most important cases of $Z = 0$, 1, and 2,
we obtain (see also Table~I of Supplementary Material \cite{suppl})
\begin{align}\label{eq:4}
G_{60}(Z = 0) = -101.1\,(5.5)\,,\nonumber \\
G_{60}(Z = 1) = -104.1\,(3.8)\,,\\
G_{60}(Z = 2) = -104.6\,(2.8)\,.\nonumber
\end{align}
Our result for $Z = 0$ differs by 3.7$\,\sigma$ from the analytical value of the $B_{60}$ coefficient,
$-61.6\,(9.2)$ \cite{pachucki:03:prl}, and for $Z = 1$,
by 3.6$\,\sigma$ from the $\Za$-expansion value,
$-66.8\,(9.6)$ \cite{pachucki:03:prl,karshenboim:19}.
At the same time, the present results are consistent with our previous, less
precise extrapolation values $G_{60}(Z = 1) = -127\,(38)$ \cite{yerokhin:05:sese}
and $G_{60}(Z = 1) = -86\,(15)$ \cite{yerokhin:09:sese}.
The previously recommended value $G_{60}(Z = 1) = -75.8\,(9.4)$ from Ref.~\cite{tiesinga:21:codata18}
was obtained by combining the $\Za$-expansion and the previous, less-accurate all-order results and deviates
from our present value by $2.8\,\sigma$.

It is important to note that the uncertainties of our extrapolated values rely on the
analytical constraints of the logarithmic coefficients in Eq.~(\ref{eq:3}). If these constraints
are revised, our uncertainties will have to be modified as well. The dominant
contributions to the uncertainties come from the $B_{71}$ and $B_{82}$ coefficients.

When speculating about possible reasons for the disagreement with the $\Za$-expansion
calculations, two plausible scenarios can be considered.
The first scenario involves an additional
contribution to the $B_{60}$ coefficient, which has
not been fully calculated yet, estimated around $\delta B_{60} \approx -40$.
The second scenario suggests a modification
in the $B_{62}$ or $B_{61}$ coefficients, which have not
been independently confirmed. Specifically, a 10\% adjustment
in the $B_{61}$ coefficient, $\delta B_{61}\approx -5$,
could restore the consistency with
the nonperturbative calculations.

Although our calculations are performed for the $1S$ state only, we are able to refine
theoretical predictions also for other $nS$ states. This is because the normalized energy differences
$\nu_{n} = n^3E(nS)-E(1S)$ are currently better understood than the individual
$nS$ energies \cite{tiesinga:21:codata18}. Consequently, the improved theoretical prediction
for the $1S$ state also enhances the accuracy of the $nS$ energies via the relation
$n^3E(nS) = \nu_n + E(1S)$.

Turning to the experimental consequences of our calculations,
the updated value of the SESE correction shifts the theoretical
prediction \cite{tiesinga:21:codata18}
for the $1S$-$2S$ Lamb shift in hydrogen by $-2.5$~kHz (2.0$\,\sigma$)
and slightly reduces its
theoretical uncertainty from $1.3$~kHz to $1.1$~kHz. This entails the fractional
shift of $-1.0 \times 10^{-12}$ in the value of the Rydberg constant.
Specifically, using the SESE contribution obtained
here, the remaining theory from Ref.~\cite{yerokhin:18:hydr}, the value of the
proton radius from Ref.~\cite{pachucki:24:rmp}, and the experimental $1S$-$2S$
transition energy from Ref.~\cite{matveev:13}, we obtain the
Rydberg constant $R_{\infty}$ multiplied by the speed of light $c$ as
\begin{align}\label{eq:Ry}
R_{\infty}c = 3\,289\,841\,960.2449\,(23)~\mbox{\rm MHz}\,,
\end{align}
which is smaller than the CODATA22 value \cite{CODATA2022} by 1.4 of
its standard deviation.

In summary, we performed calculations of the two-loop electron self-energy correction
to the $1S$ Lamb shift of hydrogen-like ions to all orders in the parameter $\Za$.
The calculational approach developed in this Letter allowed us to improve the
numerical accuracy of this effect
by more than an order of magnitude and extend calculations
to lower nuclear charges than previously possible. The improved numerical
results break the previously assumed consistency between the all-order
and the $\Za$-expansion approaches.
The extrapolation of our all-order results to hydrogen revealed a disagreement
of $3.6\,\sigma$ with the $\Za$-expansion prediction.
The obtained result for the two-loop self-energy correction for the hydrogen
is twice as precise as the currently accepted value \cite{tiesinga:21:codata18}
and differs from it by $2.8\,\sigma$.
The resulting shift in the theoretical prediction for the $1S$-$2S$ transition frequency
in hydrogen decreases the value of the Rydberg constant by $3.3$~kHz.

The revealed disagreement calls for an independent
verification of the existing $\Za$-expansion results
for logarithmic contributions of order $m\alpha^2(\Za)^6$ and for
calculations of the missing part of the $B_{60}$ coefficient.
In addition, the analytical
calculation of the $B_{71}$ coefficient is desirable, as it
would ensure the correctness of the
extrapolation of all-order results
towards $Z = 1$ and improve its accuracy.

The method developed in the present Letter opens a way for
improved calculations of other two-loop QED
corrections to the Lamb shift \cite{yerokhin:08:twoloop}
and, most importantly, the two-loop QED effects to the
bound-electron $g$ factor \cite{sikora:20}, which are
urgently needed for interpretation of existing
experimental data \cite{morgner:23}.

%
%



\begin{widetext}

%
%

\section*{Supplemental Material: Details of the subtraction scheme}

Here we describe the calculational scheme developed in the present work which allowed
us to achieve improved convergence of the partial-wave expansion. Since the
detailed description of our previous approach is available in
Refs.~\cite{yerokhin:03:epjd,yerokhin:10:sese,yerokhin:18:sese}, here we
concentrate on the new features as compared
to the previous calculational scheme. The previous scheme,
referred to as the ``standard'' one in the following,
is based on the separation of each of the
overlapping and nested diagrams in three parts:
(i) the part calculated in the coordinate space (the $M$ term), (ii) the
part calculated in the mixed momentum-coordinate representation (the $P$ term),
and (iii) the part calculated in the momentum space (the $F$ term).
In the new calculational scheme (referred to as the ``accelerated'' one in the
following), we introduce additional subtractions in the $M$ and $P$ terms
and then compute the subtracted contributions
in the closed form, in the momentum space. This leads to a re-definition of the
separation, i.e., the $M$, $P$, and $F$ terms are replaced by the $M'$, $P'$, and $F'$
terms.

In this Supplement, we will refer to formulas and figures of the main text as
Eq.~(?M) and Fig.~?M, respectively.

\subsection{Overlapping diagram}

\subsubsection{$M$ term}

In the standard scheme, the overlapping $M$ term
is obtained from Eq.~(1M) by
applying the following subtractions (as shown in the uppermost line of Fig.~2M):
\begin{align}  \label{eqM1}
 (O,M): \ \ \     G\, G\, G \  & \to
   G\, G\, G
 - G\, G^{(0)}\,G^{(0)}
 - G^{(0)}\,G^{(0)}\,G
 + G ^{(0)}\,G ^{(0)}\,G ^{(0)}
 - G ^{(0)}\,G ^{(1)}\,G ^{(0)}\,.
\end{align}
Here, the three $G$'s on the left-hand side correspond to the three Green functions in Eq.~(1M),
$G^{(0)}$ denotes the free-electron Green function, and $G^{(1)}$ denotes the one-potential Green
function,
\begin{align}
G^{(1)}(\vare,\bfx_1,\bfx_2) = \int d\bfz\, G^{(0)}(\vare,\bfx_1,\bfz)\,
V_C(z)\, G^{(0)}(\vare,\bfz,\bfx_2)\,,
\end{align}
where $V_C(x) = -\Za/x$ is the nuclear Coulomb potential. For brevity, in Eq.~(\ref{eqM1}) we omitted all arguments
of the Green functions, which can be unambiguously restored from Eq.~(1M).

In order to improve the convergence of the partial-wave expansion, we
introduce additional subtractions as follows
(see the second line from the top in Fig.~2M):
\begin{align}  \label{eqM2}
(O,M'): \ \ \     G\, G\, G \  & \to \ldots
 - 2\,V_C\, \dot{G}^{(0)} \,G^{(1)} \,G ^{(0)}
 - V_C\, G^{(0)} \,\ddot{G}^{(0)} \,G ^{(0)}\,V_C
 - V_C\, \dot{G}^{(0)} \,G^{(0)} \,\dot{G} ^{(0)}\,V_C\,,
\end{align}
where $\ldots$ corresponds to the right-hand-side of Eq.~(\ref{eqM1}),
$\dot{G} = -\partial /(\partial \vare)\,G(\vare)$ and
$\ddot{G} = (1/2)\,\partial^2 /(\partial \vare)^2\,G(\vare)$.
In Eq.~(\ref{eqM2}) and in what follows
we assume the implicit ordering of radial arguments, e.g., $V_C\,G \,G \,G \,V_C$ should be understood as
$V_C(x_1)\,G (\bfx_1,\bfx_2)\,G (\bfx_2,\bfx_3)\,G (\bfx_3,\bfx_4)\,V_C(x_4)$, etc.

\subsubsection{$P$ term}

In the standard scheme, the overlapping $P$ term takes care of the two (equivalent)
subtraction terms $G\, G^{(0)}\,G^{(0)}$ and $G^{(0)}\, G^{(0)}\,G$ in Eq.~(\ref{eqM1}). It is defined
by applying the following subtractions (see the third line from the top in Fig.~2M):
\begin{align}  \label{eqP1}
(O,P): \ \ \    G\, G^{(0)}\,G^{(0)}
\to
  G\, G^{(0)}\,G^{(0)}
  - G^{(0)}\, G^{(0)}\,G^{(0)}
  - G^{(1)}\, G^{(0)}\,G^{(0)}\,.
\end{align}
In the accelerated scheme, we
introduce an additional subtraction and replace Eq.~(\ref{eqP1}) by
(see the third line from the top in Fig.~2M)
\begin{align}  \label{eqP2}
(O,P'): \ \ \      G\, G^{(0)}\,G^{(0)}
\to
  G\, G^{(0)}\,G^{(0)}
  - G^{(0)}\, G^{(0)}\,G^{(0)}
  - G^{(1)}\, G^{(0)}\,G^{(0)}
  - V_C\,\ddot{G}^{(0)}\, G^{(0)}\,G^{(0)}\,V_C\,.
\end{align}

\subsubsection{$F$ term}

The $F$ term takes care of all subtracted terms which now contain only $G^{(0)}$ (and its
derivatives) and $G^{(1)}$, but not $G$'s.
The additional terms subtracted in the $M$ and $P$ terms are (of course)
added back and evaluated in the momentum space without any partial-wave expansion.
This leads to the diagrams shown in the bottom line of Fig.~2M.
Their calculation
is performed by the technique developed for the $F$ term in our previous investigations
\cite{yerokhin:03:epjd,yerokhin:10:sese,yerokhin:18:sese}. The calculation is long
but relatively straightforward, as all of these contributions are UV finite and
can be computed in $D = 4$ dimensions.

\subsection{Nested diagram}

\subsubsection{$M$ term}

In the standard scheme, the nested $M$ term
is obtained from Eq.~(2M) by
applying the following subtractions (see the uppermost line of Fig.~3M):
\begin{align}  \label{eqM3}
(N,M): \ \ \  G\, G\, G \  & \to
   G\, G\, G
 - G\, G^{(0)}\,G
 - G\, G^{(1)}\,G
 \,.
\end{align}
Here, we ignore the reference-state infrared (IR) divergencies present in the
nested $M$ term. Treatment of the IR divergences is the same as in our previous
studies, so we do not discuss it here.
In the accelerated scheme, we perform an additional subtraction that greatly improves the
partial-wave convergence of the nested $M$ term. Specifically, we replace Eq.~(\ref{eqM3})
by (see the uppermost line of Fig.~3M)
\begin{align}  \label{eqM4}
(N,M'): \ \ \  G\, G\, G \  & \to
   G\, G\, G
 - G\, G^{(0)}\,G
 - G\, G^{(1)}\,G
 - G\, V_C\, \ddot{G}^{(0)}\, V_C \,G
 \,.
\end{align}

\subsubsection{$P$ term}

In the standard scheme, there are two nested $P$ terms that take care of the two
subtraction terms $G\, G^{(0)}\,G$ and $G\, G^{(1)}\,G$ in Eq.~(\ref{eqM3});
they are referred to as the first and as the second, respectively.

In the standard scheme, the first nested $P$ term is defined by the
subtraction
(see the second line of Fig.~3M):
\begin{align}  \label{eqP4a}
(N1,P): \ \ \  G\, G^{(0)}\, G
  \  & \to
     G\, G^{(0)}\, G
 -    G^{(0)}\, G^{(0)}\, G^{(0)}
 -    2\,G^{(1)}\, G^{(0)}\, G^{(0)}
 \,.
\end{align}
In the accelerated scheme, we introduce additional subtractions that improve
the partial-wave expansion (see the second line of Fig.~3M):
\begin{align}  \label{eqP4b}
(N1,P'): \ \ \  G\, G^{(0)}\, G
  \  & \to
\ldots
 -    V_C\,\dot{G}^{(0)}\, G^{(0)}\, \dot{G}^{(0)}\, V_C
 -    2\,V_C\,\ddot{G}^{(0)}\, G^{(0)}\, G^{(0)}\, V_C
 \,,
\end{align}
where $\ldots$ stand for the right-hand-side of Eq.~(\ref{eqP4a}).

In the standard scheme, the second nested $P$ term is defined by the
subtraction
(see the third line of Fig.~3M):
\begin{align}  \label{eqP5a}
(N2,P): \ \ \  G\, G^{(1)}\, G
  \  & \to
     G\, G^{(1)}\, G
 -    G^{(0)}\, G^{(1)}\, G^{(0)}
 \,.
\end{align}
In the accelerated scheme, we introduce an additional subtraction that improves
the partial-wave expansion (see the third line of Fig.~3M):
\begin{align}  \label{eqP5b}
(N2,P'): \ \ \  G\, G^{(1)}\, G
  \  & \to
\ldots
 -    2\,V_C\,\dot{G}^{(0)}\, \dot{G}^{(0)}\, G^{(0)}\, V_C
 \,,
\end{align}
where $\ldots$ stand for the right-hand-side of Eq.~(\ref{eqP5a}).

In addition, in the accelerated scheme we introduce the third nested $P$ term that
takes care of the additional subtraction in Eq.~(\ref{eqM4})
(see the fourth line of Fig.~3M):
\begin{align}  \label{eqP5c}
(N3,P'): \ \ \  G\, V_C\, \ddot{G}^{(0)}\, V_C \,G
  \  & \to
G\, V_C\, \ddot{G}^{(0)}\, V_C \,G
-
V_C\, G^{(0)}\, \ddot{G}^{(0)}\, G^{(0)} \,V_C
 \,.
\end{align}
This contribution contains both the Dirac-Coulomb Green function $G$ and the UV
diverging free self-energy operator. Its calculation is very similar to that of
the first nested term.

\subsubsection{$F$ term}

Similarly to the overlapping $F$ term, the nested $F$ term takes care of all subtracted
term in the nested $M$ and $P$ terms. The corresponding diagrams are shown in the
two bottom lines of Fig.~3M.
Their calculation
is performed by the technique developed for the $F$ term in our previous investigations
\cite{yerokhin:03:epjd,yerokhin:10:sese,yerokhin:18:sese}.

\end{widetext}

%
%

\section*{Supplemental Material: Details of the extrapolation procedure}

We now describe the details of our extrapolation of the all-order numerical results for
the higher-order remainder $G_{60}(\Za)$ summarized in Table~I of the main text to lower values of
$Z$. The analytical form of the $\Za$ expansion of $G_{60}(\Za)$ is given by Eq.~(5) of the
main text.
The main difficulty of the extrapolation arises from the presence of logarithmic terms in the
$\Za$ expansion. The problem is that
we cannot constraint the logarithmic coefficients based solely on the available numerical data.
This is because the unknown logarithmic terms, such as $(\Za)\ln (\Za)^{-2}$, exhibit a distinct
structure only for very small values $Z$, whereas for $Z>5$, where the numerical data is available,
they are smooth functions that can be reasonably approximated by polynomials. As a result,
if one uses a fitting function with logarithmic terms, the fitting minimization
procedure does not have a well-localized minimum, and the resulting logarithmic coefficients
cannot be obtained unambiguously.
For this reason we choose to
approximate the numerical data with polynomials only and incorporate the effects of missing
logarithmic terms into the estimation of uncertainty, using the existing constraints
derived from the $\Za$ expansion.

Our extrapolation procedure is as follows. First, we subtract all known logarithmic contributions
from the higher-order remainder
$G_{60}(\Za)$,
\begin{align}
G^{\prime}_{60}(\Za) = &\ G_{60}(\Za) - B_{72}\,(\Za)\,L^2
 \nonumber \\ &
- B_{71}\,(\Za)\,L - B_{84}\,(\Za)^2\,L^4\,,
\end{align}
where for $B_{71}$ we take the central value of the analytical estimate from Eq.~(6) of the
main text.
We then extrapolate the numerical data for $G^{\prime}_{60}$ and later
re-add the subtracted logarithmic
terms to obtain $G_{60}$.

Our extrapolation is performed into two steps. First, we approximate the numerical data for $G^{\prime}_{60}(Z)$
with polynomials $P_n(Z)$ of degree $n = 2$, 3, and 4 and take the
$P_3(Z)$ approximation as the central value. The uncertainty of the polynomial
extrapolation is estimated as twice the maximal difference between the three fitting functions.
In the second step, we evaluate additional uncertainties due to the presence of logarithmic terms.
To do this, we subtract
each of the unknown logarithmic terms in Eq.~(5) of the main text from $G^{\prime}_{60}$,
with coefficient given by the upper  (lower) limit of the corresponding constraint in Eq.~(6) of the main text.
We then repeat the cubic extrapolation of the subtracted numerical data and then re-add the
subtracted logarithmic term back. The difference with
the unsubtracted cubic extrapolation
gives us the estimate of the uncertainty due to this logarithmic term. We repeat this procedure
for each of the unknown logarithmic terms and combine all uncertainties -- one due to
the polynomial extrapolation and others due to four unknown logarithmic terms --
quadratically.

The polynomial extrapolation is carried out by minimizing the following functional
\begin{align}
\chi^2 = \sum_i \frac{[G^{\prime}_{60}(Z_i) - P_n(Z_i)]^2}{w_i^2}\,,
\end{align}
where $Z_i = 8,\ldots,50$ and $w_i$ are weights. The data points with $Z = 5$, 6,
and 7 were excluded because of their large uncertainty.
Inspired by Ref.~\cite{karshenboim:19:sese}, we take the weights $w_i^2$ as the quadratic sum
of the numerical error of $i$th data point and an additional effective error arising from
the presence of terms scaling as $Z^k$ with $k > n$ in the numerical data.
Specifically, we take
\begin{align}
w_i^2 = \delta_i^2 + (a\,Z_i^{n+1})^2\,,
\end{align}
where $\delta_i$ is the numerical uncertainty of the data point $i$, $n$ is the degree of
the fitting polynomial, and $a$ is some coefficient. The value of $a$ is
selected by requiring that variations of the fitting result at $Z = 0$
caused by removal of the first two and the last two data points are small and approximately
equal.

Results of our extrapolation for the higher-order remainder
$G_{60}$ and its uncertainty $\delta G_{60}$ are presented in Table~\ref{tab:fit}. The main
sources of uncertainty are (i) the coefficient $B_{71}$,
(ii) the coefficient $B_{82}$,
(iii) the polynomial extrapolation.  E.g., for $Z = 0$, the corresponding uncertainty
contributions are 3.9, 2.7, and 2.4.

\begin{table}[b]
    \caption{Fitting results for the
    two-loop self-energy higher-order remainder $G_{60}$ and its uncertainty $\delta G_{60}$.
    \label{tab:fit}}
\begin{ruledtabular}
    \begin{tabular}{l w{5.5} w{5.5} }
 \multicolumn{1}{c}{$Z$}
    & \multicolumn{1}{c}{$G_{60}(Z)$}
    & \multicolumn{1}{c}{$\delta G_{60}(Z)$} \\
\hline\\[-5pt]
 0   &  -101.1370   &     5.5000 \\
 1   &  -104.0993   &     3.8027 \\
 2   &  -104.6048   &     2.8096 \\
 3   &  -104.4381   &     2.0909 \\
 4   &  -103.8999   &     1.5448 \\
 5   &  -103.1215   &     1.1219 \\
 6   &  -102.1758   &     0.7919 \\
 7   &  -101.1087   &     0.5330 \\
 8   &   -99.9510   &     0.3315 \\
    \end{tabular}
\end{ruledtabular}
\end{table}

\end{document}